\documentclass[fleqn,12pt,twoside]{article}
\usepackage{espcrc1}

\usepackage{graphicx,epsf}
\usepackage[figuresright]{rotating}

\newcommand{\AmS}{{\protect\the\textfont2
  A\kern-.1667em\lower.5ex\hbox{M}\kern-.125emS}}

\title{Deep Virtual Compton Scattering and the Nucleon Generalized Parton 
Distributions.}

\author{M. Guidal\address{Institut de Physique 
	Nucl\'eaire, IN2P3, F-91406 Orsay, France, \\ 
        E-mail: guidal@ipno.in2p3.fr}}
       
\begin{document}

\maketitle

\begin{abstract}
In the following, the subject of Deep Virtual Compton Scattering on the 
nucleon and its relation to the recently introduced concept of Generalized 
Parton Distributions are briefly reviewed. The general theoretical framework and
the links between theory and experiment will be outlined and the recently 
published data which look promising for the development of this field will
be discussed. Finally, the experimental prospectives of the domain will be presented.
\end{abstract}

\section{Introduction}

The scattering of coherent light on an object (broadly speaking, Compton scattering) 
is one of the most elementary processes of physics. In a general way,
by measuring the angular and energy distributions of the scattered 
light, information on the structure and the shape of the 
probed object can be accessed : for instance, the spatial or momentum distributions 
of its internal constituents, their spin distributions, etc... While such scattering 
can certainly be done
with other incident particles, the advantageous feature
of light (be it under the form of a wave or a photon particle)
is its electromagnetic nature, which makes it interact 
with matter through the most precise theory we know, Quantum
Electrodynamics (QED) at the most fundamental level. 

The wavelength (inversely proportional to the energy) of the incident 
light must match the size of the probed object in order to be able
to probe its internal structure. Over time, the 
scattering of light has been widely studied and used to probe objects of decreasing 
size (and therefore of increasing energy). 
In order to study the quark and gluon structure of the nucleon (i.e., the partons), 
incident beams on the GeV scale are needed to probe sizes of the order of a 
fermi. It is only recently, with the advent of intense multi-GeV 
lepton beam facilities, that it has become possible to experimentally study
Compton scattering at the smallest dimensions of matter~: the nucleon or quark 
and gluon level, 
where it is traditionally called Deep Virtual Compton Scattering (DVCS)
-the term ``virtual" here meaning that the incoming photon is radiated from a lepton beam, 
which presents the additional advantage of varying its 3-momentum independently 
of its energy-.

In parallel, only less than 10 years ago, the theoretical formalism has appeared,
within the framework of QCD (``Quantum ChromoDynamics", the theory governing the interaction
between quarks and gluons), to interpret such reaction at the
partonic level, through the concept of the ``Generalized Parton Distributions"
(GPDs). In the following, the general theory and experimental status of DVCS,
that is the Compton scattering at the nucleon level, shall be briefly presented.

\section{Generalized Parton Distributions}
\label{sec1}

\subsection{Formalism}

\begin{figure}[ht]
\epsfxsize=11 cm
\epsfysize=6. cm
\centerline{\epsffile{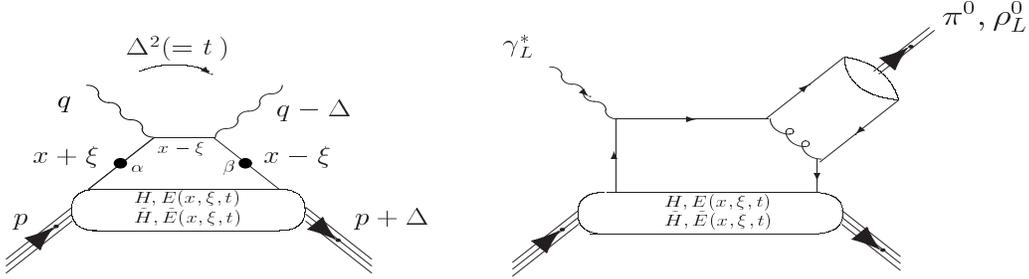}}
\vspace{-1.5cm}
\caption{``Handbag" diagrams~: a) for DVCS (left) and b) for meson 
production (right).}
\label{fig:handbags}
\end{figure}

In the last decade, Mueller et al.~\cite{Mu94}, Ji~\cite{Ji97} and 
Radyushkin~\cite{Rady} have shown that the leading order perturbative
QCD (pQCD) amplitude for Deeply Virtual Compton Scattering in
the forward direction can be factorized, in the Bjorken regime
(i.e., in simplifying, large $Q^2$, where -$Q^2$ is
the squared mass of the virtual photon)
in a hard scattering part, exactly calculable in pQCD or QED,
and a nonperturbative nucleon structure part. This is illustrated in 
Fig.~\ref{fig:handbags}a). In these so-called ``handbag" diagrams, 
the lower blob represents the unknown structure of the nucleon 
and can be parametrized, at leading order pQCD, in 
terms of 4 generalized structure functions, the GPDs. Using Ji's notation, 
these are called $H, \tilde H, E, \tilde E$, and  
depend upon three variables~: $x$, $\xi$ and $t$. One considers the
process in a frame where the proton has a large momentum along
a certain direction which defines the longitudinal components.
 
$x+\xi$ is the longitudinal
momentum fraction carried by the initial quark struck by the incoming virtual photon.
Similarly, $x-\xi$ relates to the final quark going back in the nucleon
after radiating the outgoing photon. The difference
in the longitudinal momentum fraction between the initial and final quarks
is therefore $-2\xi$. 
In comparison to $-2\xi$ which refers to purely {\it longitudinal}
degrees of freedom, $t$, the squared 4-momentum transfer between 
the final nucleon and the initial one, contains {\it transverse} degrees 
of freedom (so-called ``$k_\perp$") as well.

Intuitively, the GPDs represent the probability amplitude of finding a
quark in the nucleon with a longitudinal momentum fraction $x-\xi$
and of putting it back into the nucleon with a longitudinal momentum
fraction $x+\xi$ plus some transverse momentum ``kick", which is
represented by $t$. Explicitly, the matrix element of the bilocal quark operator,
representing the lower blob in Figs.~\ref{fig:handbags}a) 
and~\ref{fig:handbags}b), reads at leading twist~: 
\begin{eqnarray}
&&{{P^+} \over {2 \pi}}\, \int d y^- e^{i x P^+ y^-} 
\langle p^{'} | \bar \psi_\beta (- {y \over 2}) \psi_\alpha ({y \over 2}) 
| p \rangle  {\Bigg |}_{y^+ = \vec y_{\perp} = 0} \nonumber\\
&&= {1 \over 4} \left\{ ({\gamma^-})_{\alpha \beta} 
\left[ {H^q(x,\xi,t) \; \bar N(p^{'}) \gamma^+ N(p) 
\;+\; E^q(x,\xi,t) \; \bar N(p^{'}) i \sigma^{+ \kappa} 
{{\Delta_{\kappa}} \over {2 m_N}} N(p)} \right] \right. \nonumber\\ 
&&\hspace{.7cm}\left. + ({\gamma_5 \gamma^-})_{\alpha \beta} 
\left[ {\tilde H^q(x,\xi,t) \; \bar N(p^{'}) \gamma^+ \gamma_5 N(p) 
\;+\; \tilde E^q(x,\xi,t) \; \bar N(p^{'}) \gamma_5 
{{\Delta^+} \over{2 m_N}} N(p) } \right] \right\}\;,
\label{eq:qsplitting}
\end{eqnarray}
where $\psi$ is the quark field, $N$ the nucleon spinor and 
$m_N$ the nucleon mass. One uses a frame where the virtual photon 
momentum $q^\mu$ and the average nucleon momentum $P^\mu$ are collinear
along the $z$-axis and in opposite directions. 

The GPDs actually reveal a ``double" nature~: 
since negative momentum fractions are identified with with antiquarks, one can define two 
regions according to whether $|x| > \xi$ or $|x| < \xi$. In the region 
$ -\xi < x < \xi$, 
one ``leg" in Fig.~\ref{fig:handbags}a), represents a quark, and the other an 
antiquark. In this region, the GPDs behave like a meson distribution amplitude
and can be interpreted as the probability amplitude of finding 
a quark-antiquark pair in the nucleon. This kind of information on
$q{\bar q}$ configurations in the nucleon and, more generally, the
correlations between quarks (or antiquarks) of different momenta,
all, being information carried in the concept of GPDs, are
completely unknown at the time being, and reveal the richness and novelty of
the GPDs. 

In Eq.(\ref{eq:qsplitting}), one can see that $H$ and $E$ are independent 
of the quark helicity and are therefore called {\it unpolarized}
GPDs, whereas $\tilde H$ and $\tilde E$ are helicity dependent and are
called {\it polarized} GPDs. The GPD's $H^q, E^q, \tilde H^q, \tilde E^q$ 
are defined for a single quark flavor ($q = u, d$ and $s$). The GPDs $H$ and $\tilde H$ 
are actually a generalization of the parton distributions 
measured in deep inelastic scattering. Indeed, in the forward 
direction, $H$ reduces to the quark distribution and $\tilde H$ to the 
quark helicity distribution measured in deep inelastic scattering : 

\begin{equation}
H^q(x,0,0)\,=\, q(x) \;,\hspace{2cm}
\tilde{H}^q(x,0,0) \,=\, \Delta q(x) \;,
\end{equation}

Furthermore, at finite momentum transfer, there are
model independent sum rules which relate 
the first moments of these GPDs to the elastic form factors.
By integrating Eq.(\ref{eq:qsplitting}) over $x$, one gets the following
relations for one quark flavor :
\begin{eqnarray}
\label{eq:vecsumrule}
&&\int_{-1}^{+1} d x H^{q}(x,\xi,t) \,=\, F_1^{q}(t) \;, \hspace{0.5cm}
\int_{-1}^{+1} d x E^{q}(x,\xi,t) \,=\, F_2^{q}(t) \;, \\
&&\int_{-1}^{+1} d x \tilde H^{q}(x,\xi,t) \,=\, g_A^{q}(t) \;, \hspace{0.5cm}
\int_{-1}^{+1} d x \tilde E^{q}(x,\xi,t) \,=\, h_A^{q}(t) \;.
\label{eq:axvecsumrule}
\end{eqnarray} 
where $F_1$ and $F_2$ are related to the nucleon electromagnetic 
form factors and $g_A$ and $h_A$ denote the axial and pseudoscalar form 
factors of the nucleon.

It has been shown~\cite{Burkardt:2000,Diehl:2002he,pire} 
that the $t$ dependence of the GPDs can be related, via a Fourier 
transform, to the transverse spatial distribution of the partons in the 
nucleon. At $\xi$=0, the GPD($x,0,t$) can be interpreted as
the probability 
amplitude of finding in a nucleon a parton with {\it longitudinal} momentum 
fraction $x$ at a given {\it transverse} impact parameter, related to $t$.
In this way, the information contained in a traditionnal
parton distribution, as measured in inclusive Deep Inelastic Scattering (DIS),
and that contained within a form factor, as measured in 
elastic lepton-nucleon scattering, are now combined and correlated
in the GPD description~\cite{Belitsky:2003nz}.

The second moment of the GPDs is relevant to the nucleon spin structure. 
It was shown in Ref.\cite{Ji97} that there exists a (color) gauge-invariant 
decomposition of the nucleon spin:
\begin{equation}
{1 \over 2} \,=\, J_q \,+\, J_g \;,
\end{equation}
where $J_q$ and $J_g$ are respectively the total quark and gluon spin
 contributions to the nucleon total angular momentum. 
The second moment of the GPD's gives 
\begin{equation}
J_q \,=\, {1 \over 2} \, \int_{-1}^{+1} d x \, x \, 
\left[ H^{q}(x,\xi,t = 0) + E^{q}(x,\xi,t = 0) \right] \;, 
\label{eq:dvcs_spin}
\end{equation}
and this relation is independent of $\xi$. 
The total quark spin contribution $J_q$ decomposes as 
\begin{equation}
J_q = {1 \over 2} \Delta \Sigma + L_q \;,
\label{eq:spindecomp}
\end{equation}
where 1/2 $\Delta \Sigma$ and $L_q$ are respectively 
the quark spin and quark orbital contributions to the nucleon spin. 
Since $\Delta \Sigma$ has been measured through polarized DIS experiments
(it is about 20\%) and $J_g$ is currently being measured at COMPASS and RHIC,
a measurement of the sum rule of Eq.(\ref{eq:dvcs_spin}) in terms of
the GPD's provides a model independent way of determining the quark orbital 
contribution to the nucleon spin, and therefore complete the ``spin-puzzle".

The GPDs reflect the structure of the nucleon 
independently of the reaction which probes the nucleon. They can also be 
accessed through the hard exclusive electroproduction of mesons 
-$\pi^0$, $\rho^0$, $\omega$, $\phi$, etc.- 
(see Fig.~\ref{fig:handbags}b)) for which a QCD factorization
proof was given in Ref.~\cite{Collins97}. In this case, the
perturbative part of the diagram involves a 1-gluon exchange and 
therefore the strong running coupling constant, whose behavior
at low energy scales is not fully controlled, makes
the calculations and the interpretation of the data more 
complicated.

The current theoretical activity in the field bears mainly on 
the modelization of the GPDs within different frameworks (to name a few, 
the chiral quark soliton 
model~\cite{petrov}, the constituent quark model~\cite{vento,pasqui}, 
light-cone wavefunction overlap~\cite{feldmann},...),
on the control of the QCD corrections 
(Next to Leading Order evolution~\cite{evo}, higher twists~\cite{freundht,kivel}, 
....), on lattice
calculations~\cite{:2003is,qcdsf} and the extension of the formalism to 
processes other than the ``hard" electroproduction of photons and mesons on the nucleon.

A non-exhaustive list of currently explored reactions is~: 
$\gamma p\to\gamma^* p$~\cite{berger} where $\gamma^*$ decays into
a lepton pair ({\it Timelike} Compton Scattering),
$\gamma^* p\to\gamma^* p$~\cite{Gui03,bel03} ({\it Double} Deep Virtual Compton 
Scattering),
$\gamma^* p\to\gamma \Delta$~\cite{deltavcs} ({\it Resonant} Deep Virtual Compton 
Scattering),
$\gamma^* A\to\gamma A$~\cite{kirch,polynuc,scop04,cano} ({\it Nuclear} Deep Virtual Compton 
Scattering), 
$\gamma p\to\gamma p$~\cite{rady98,diehl99} ({\it Real} Compton Scattering),
$\bar p p\to\gamma \gamma$~\cite{idvcs} ({\it Inverse} Compton Scattering)
-however, for these 2 last processes, RCS and ICS, the factorisation proof 
still remains to be established-, ``hard" hybrid electroproduction~\cite{anikin}, 
``hard" pentaquark electroproduction~\cite{szym}, etc...

We refer to Refs.~\cite{goeke,die} for very complete recent reviews 
of the field and more details on all these aspects which cannot 
be covered in this short contribution.

\subsection{From theory to experiment}
\label{thexp}

As mentioned in the previous section, the GPDs depend on three 
variables~: $x$, $\xi$ and $t$.
However, it has to be realized that only two of these three variables
are accessible experimentally, i.e. $\xi$ (fully
defined by detecting the scattered lepton $\xi=\frac{x_B}{2-x_B}$, where
$x_B$ is the traditional Bjorken variable used in DIS) 
and $t$ (fully defined by detecting either the recoil
proton or the outgoing photon or meson). However, $x$ (which is different 
from $x_B$ !) is a  variable which is integrated over, due to the loop in the handbag 
diagrams (see Fig.~\ref{fig:handbags}).
This means that, in general, a differential cross section will be
proportional to~:
$\mid \int_{-1}^{+1}d x {{H(x,\xi,t)} \over {x - \xi + i \epsilon}}+...\mid^2$
(where ... stands for similar terms for $E$, $\tilde{H}$, $\tilde{E}$ ;
${1} \over {x - \xi + i \epsilon}$ is the propagator of the quark 
between the incoming virtual photon and the outgoing photon -or meson-, see 
Fig.~\ref{fig:handbags}).
In general, one will therefore measure integrals (with a propagator as 
a weighting function) of GPDs. 

\begin{figure}[htb]
\epsfxsize=8. cm
\epsfysize=8. cm
\centerline{\epsffile{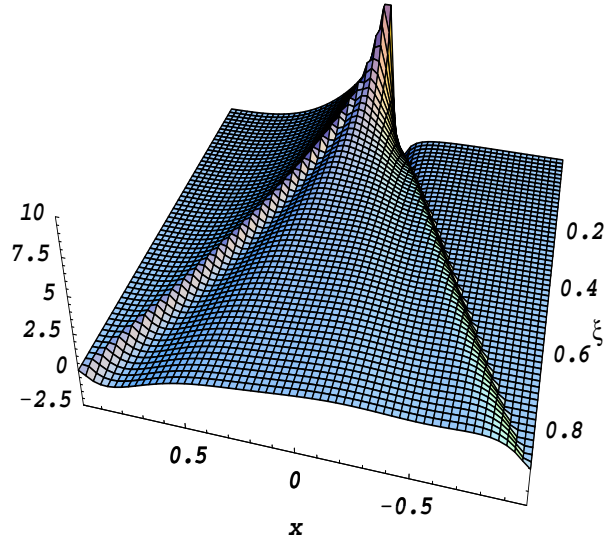}}
\vspace{-1.5cm}
\caption{One model~\cite{goeke,marcprl} for the GPD $H$ as a function of $x$ and $\xi$ for $t$=0. 
One recognizes for $\xi$=0 the typical shape of a parton distribution (with 
the sea quarks rising as $x$ goes to 0, the negative $x$ part being interpreted 
as the antiquark contribution). The figure is taken from Ref.~\cite{goeke}.}
\label{fig:spd}
\end{figure}
 
To illustrate this point, Fig.~\ref{fig:spd} shows one particular 
model~\cite{goeke,marcprl} 
for the GPD $H$ as a function of $x$ and $\xi$ (at $t=0$). One identifies at
$\xi=0$ a standard quark density distribution, with the rise
around $x=0$ corresponding to the diverging sea contribution.
The negative $x$ part is related to antiquarks. One sees that the
evolution with $\xi$ is not trivial and that measuring the integral over $x$
of a GPD, at constant $\xi$, will not define it uniquely .

A particular exception is when one measures an observable 
sensitive to the {\it imaginary} part of the amplitude, for instance, the
beam spin asymmetry -BSA- in DVCS. It is non-zero at leading order due to the
interference with the Bethe-Heitler process (see section~\ref{sec3}). Then, 
since the amplitude $\int_{-1}^{+1}d x {H(x,\xi,t) \over {x - \xi + i \epsilon}}=
PP(\int_{-1}^{+1}d x {H(x,\xi,t) \over {x - \xi}})-i\pi H(\xi,\xi,t)$, one 
actually measures the GPDs directly at some specific point, $x=\xi$
(i.e., $H(\xi,\xi,t)$). Consequently, measuring an observable that is
sensitive to the real part of the amplitude (for instance, the beam charge 
asymmetry for DVCS) gives access to $\int_{-1}^{+1}d x {H(x,\xi,t) \over {x - \xi}}$.

For mesons, transverse target polarization observables are also sensitive to a 
different combination of GPDs, namely combinations of the type~: $\int_{-1}^{+1}d x {H(x,\xi,t) \over {x - \xi}}\times
E(\xi,\xi,t)$. The exact formula is more complicated, see for 
instance Ref.~\cite{deltavcs,goeke}. Such transverse spin asymmetries 
are sensitive to a {\it product} of the GPDs, as opposed to a sum of their
squares, as is the case for a typical differential cross section.

It will therefore be a non-trivial task 
to actually extract the GPDs from the experimental observables as 
one actually only accesses, in general, weighted integrals of GPDs, or 
GPDs at some very specific points, or the product of these two. In the absence of 
any model-independent ``deconvolution" procedure at this time, one has to rely 
on some global model fitting procedure.

As previously mentioned, GPDs are defined for one quark flavor 
$q$ (i.e. $H^q$, $E^q$,...) in a way similar to standard quark distributions. This 
flavor separation can be done through the measurement of several isospin channels ;
for example, $\rho^0$ production is proportional to 
$2/3 H^u + 1/3 H^d$ (in a succinct notation) while $\omega$ production is proportional to 
$2/3 H^u - 1/3 H^d$. Similar arguments apply to the polarized GPDs with
the $\pi^{0,\pm}$, $\eta$,... channels. Similarly, DVCS on the proton
and on the neutron probe different flavor combinations of the GPDs.

In summary, a full experimental program aiming
at the extraction of the individual GPDs is a broad project which requires 
the study of several isospin channels and several observables, each 
having its own characteristics. Only a \underline{global} study
and fit to all this information may allow an actual extraction of the GPDs.

\section{Experimental aspects}
\label{sec2}

Over the last 20 years, most of what we have learned on the structure
of the nucleon has come from the \underline{inclusive}
scattering of high energy leptons on the nucleon.
By detecting only the scattered electron,  
a tremendous amount of information has already been obtained~: 
apart from having shown the quark and gluon substructure of the nucleon, these
experiments have shown that, for instance, about half of its momentum is carried by 
the quarks (the other half being carried by gluons) and that, as has been 
mentioned earlier, no more than about 20\% of the spin of the nucleon 
originates from the quarks' intrinsic spin.

Processes such as DVCS (or more generally meson leptoproduction
reactions) require the determination of a particular final
state~: not only must the scattered electron be detected but
also the whole final state. This is termed as  
an \underline{exclusive} reaction. 

The advent of the new generation of high-energy,
high-luminosity lepton accelerators combined with large
acceptance and high resolution spectrometers has recently given rise 
to the possibility of unambiguously measuring these exclusive low
cross-section processes. We now discuss the 
first experimental results which have emerged in recent
years and the new prospectives opening up.

\subsection{Recent experimental results}
\label{sec3}

\begin{figure}[htb]
\begin{minipage}[t]{76mm}
\epsfxsize=5.5 cm
\epsfysize=9. cm
\epsffile{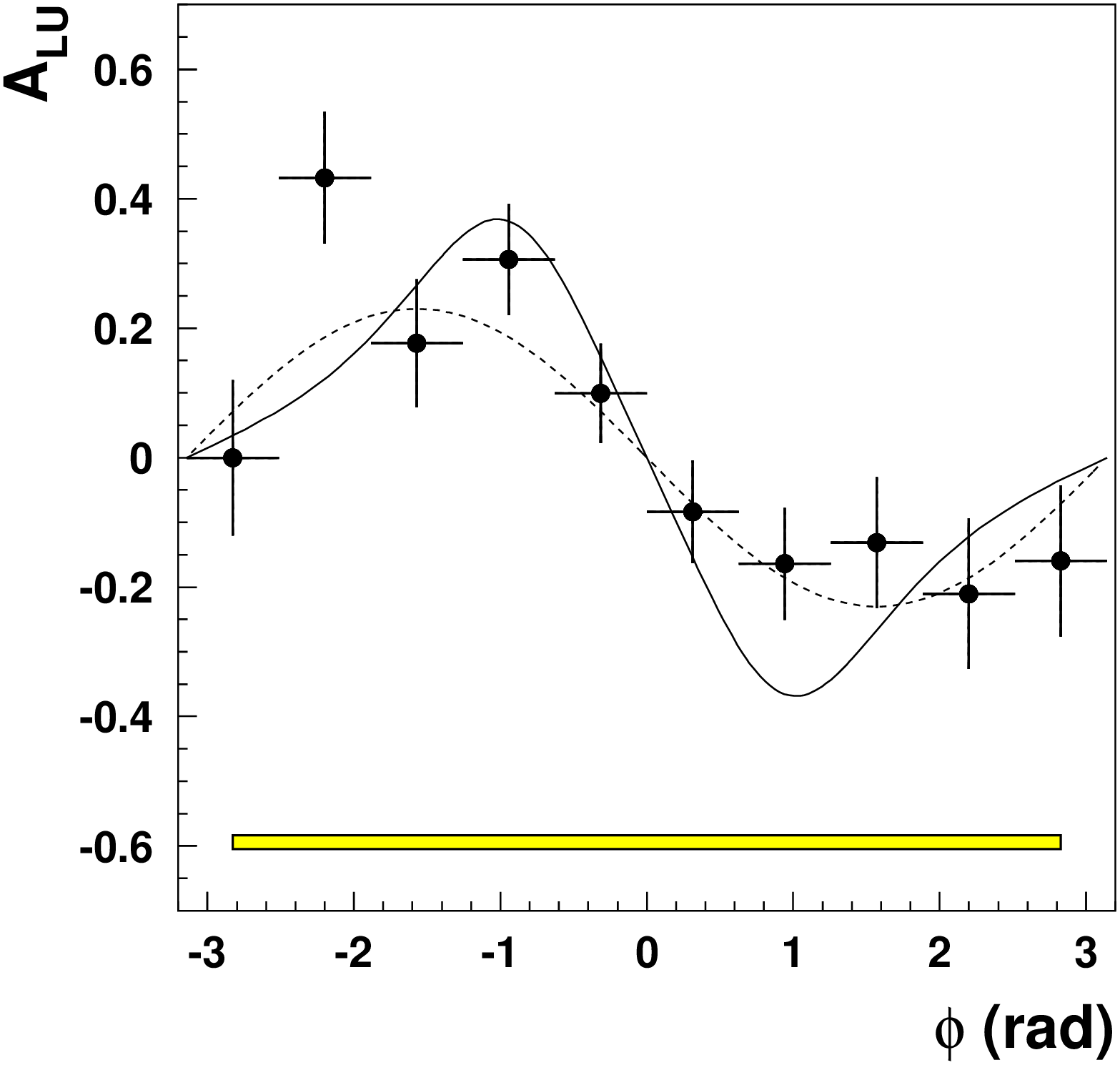}
\vspace{-2.8cm}
\caption{The DVCS beam asymmetry as a function of the azimuthal
angle $\Phi$ as measured by HERMES~\cite{dvcshermes}. Average kinematics 
are~: $<x_B>$=.11, $<Q^2>$=2.6 GeV$^2$ and $<-t>$=.27 GeV$^2$. The dashed curve
is a sin$\Phi$ fit whereas the solid curve is the theoretical GPD calculation
of Ref.~\cite{kivel}.}
\label{fig:hermes_dvcs}
\end{minipage}
\hspace{\fill}
\begin{minipage}[t]{78mm}
\epsfxsize=5.5 cm
\epsfysize=6.45 cm
\epsffile{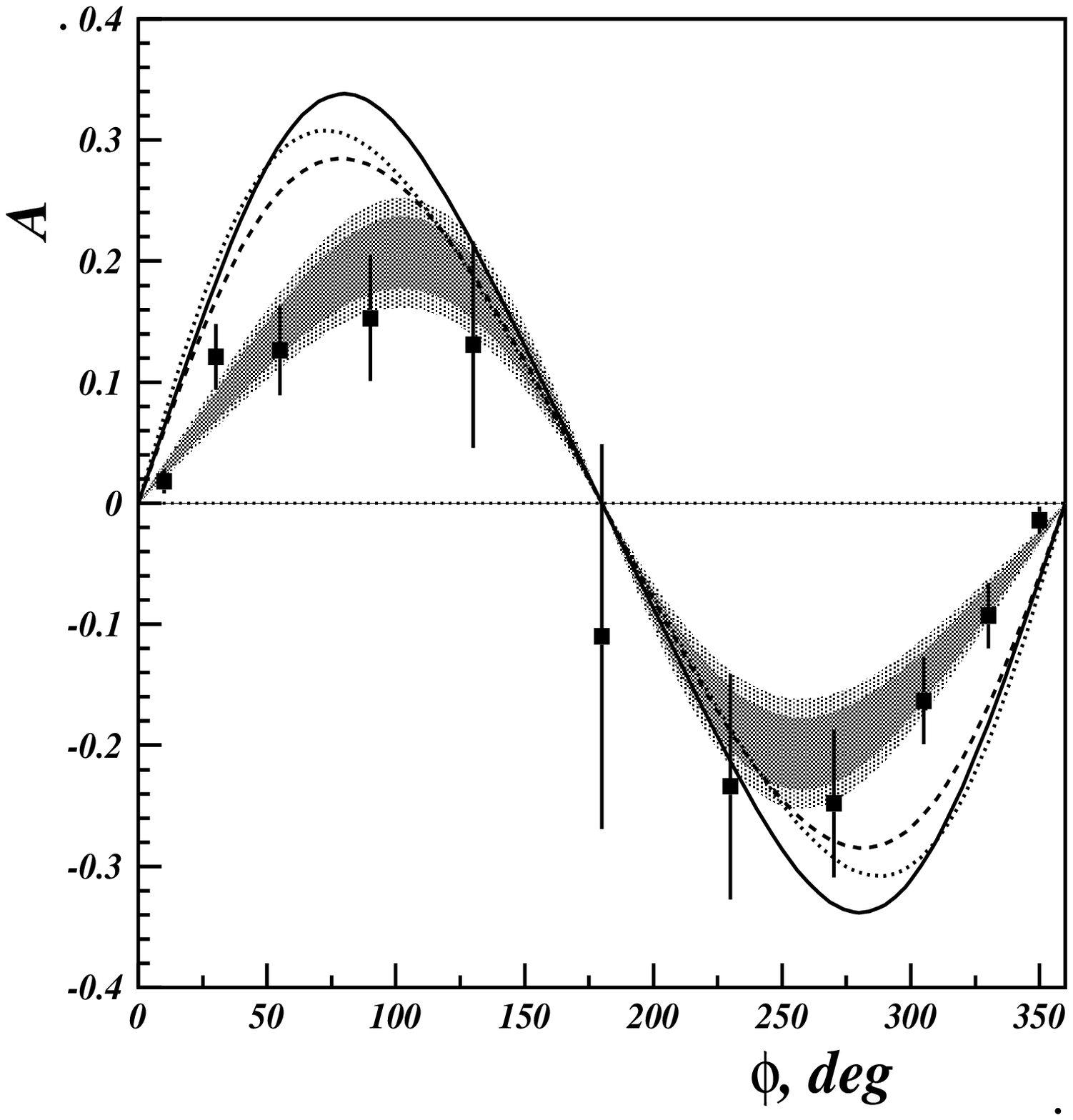}
\vspace{-2.8cm}
\caption{The DVCS beam asymmetry as a function of the azimuthal
angle $\Phi$ as measured by CLAS~\cite{dvcsclas}. Average
kinematics are~: $<x_B>$=.19, $<Q^2>$=1.25 GeV$^2$ and $<-t>$=.19 GeV$^2$.
The shaded regions are error ranges for sin$\Phi$ and sin$2\Phi$ fits.
Calculations are~: leading twist {\it without} $\xi$ dependence~\cite{vcsrev,marcprl} (dashed curve), 
leading twist {\it with} $\xi$ dependence~\cite{vcsrev,marcprl}
(dotted curve) and leading twist + twist-3~\cite{kivel} (solid curve).}
\label{fig:clas_dvcs}
\end{minipage}
\end{figure}
 
It is only in the past 3 years that some experimental results relevant to GPD physics,
and of sufficient precision, have been obtained. 

Fig.~\ref{fig:hermes_dvcs} shows the first measurement of the
BSA for DVCS on the proton by HERMES with a 27 GeV positron beam. This asymmetry
arises from the interference of the ``pure" DVCS process (where the outgoing
photon is emitted by the nucleon) and the Bethe-Heitler (BH) process (where
the outgoing photon is radiated by the incoming or scattered lepton).
The two processes are experimentally indistinguishable and interfere.
The cross section will therefore be proportionnal to the squared
amplitude~: $\mid M_{DVCS} + M_{BH} \mid^2$ while the difference between 
cross sections for different beam helicities will be proportional
to~: $\mid M_{DVCS} \times M_{BH} \mid$.
The BH process being exactly calculable in QED, this latter observable therefore gives 
access in a {\it linear} fashion to the GPDs. The difference of 
cross sections for different beam helicities is also sensitive to the
imaginary part of the amplitude~: the BH being purely real, the GPD
is measured in this way at the kinematical point 
($x=\xi,\xi,t$) as mentioned in a previous section.

The beam asymmetry, which is this latter difference of cross sections
divided by their sum, is more straightforward to access experimentally since 
normalization and systematic issues cancel, in a first order approximation, 
in the ratio. For this asymmetry, a shape close to sin$\Phi$ (not an exact sin$\Phi$ 
shape as higher twists and the Bethe Heitler have some more complex
$\Phi$ dependence) is expected, where
$\Phi$ is the angle between the leptonic and the hadronic plane. 
At HERMES, the average kinematics 
is $<x_B>$=.11 ($x_B$ is related to $\xi$, see section~\ref{thexp}), 
$<Q^2>$=2.6 GeV$^2$ and $<-t>$=.27 GeV$^2$ for which an 
amplitude of .23 for the sin$\Phi$ moment is extracted from the 
fit~\cite{dvcshermes}. 
The discrepancy between the theoretical prediction and the data in 
Fig.~\ref{fig:hermes_dvcs} can certainly be attributed, in part, to 
the large kinematical range over which the experimental data have been 
integrated, and where the model can vary significantly, but also 
to higher twist corrections not yet calculated (so far, only twist-3 corrections are under 
theoretical control for this process, the leading twist being twist-2). 
See for instance Refs.~\cite{kivel,anikin2,bel2} for more discussion on 
twist-3 accuracy in the DVCS process and Refs~\cite{diehlBSA,belBSA} about 
beam spin asymmetry in general.

The DVCS reaction at HERMES is identified by detecting the scattered
lepton (\underline{positron}) and the outgoing photon from which the missing 
mass of the non-detected proton is calculated. Due to the limited resolution of 
the HERMES detector, the selected peak around the proton mass is 
$-1.5<M_X<1.7$ GeV, which means that contributions to this asymmetry 
from nucleon resonant states as well, cannot be excluded.

This same observable, i.e. the DVCS BSA on the proton, has been measured at JLab 
with a 4.2 GeV \underline{electron} beam and the 4$\pi$ CLAS detector~\cite{dvcsclas}. 
Due to the lower
beam energy compared to HERMES, the kinematical range accessed at JLab is different~: 
$<~x_B~>$=.19, $<Q^2>$=1.25 GeV$^2$ and $<-t>$=.19 GeV$^2$. 
In this case, the DVCS reaction was identified by detecting 
the scattered lepton and the recoil proton. The missing mass of the photon
was then calculated. Due to the geometry of the CLAS
detector, outgoing photons emitted at very forward angles
escape detection. The contamination by $ep\to ep\pi^0$ events
can be estimated to some extent, and subtracted bin per bin, resulting in a 
relatively clean signature of the exclusivity of the reaction.

Figure~\ref{fig:clas_dvcs} shows the CLAS measured asymmetry
along with theoretical calculations (predictions). They are in fair agreement.
The different sign of the CLAS BSA relative to HERMES is due to the
use of electron beams in the former case compared to positron beams in
the latter. 
Again, discrepancies can be assigned to the fact that the theory is 
calculated at a single, well-defined, kinematical point whereas data have 
been integrated over several variables and wide ranges. Furthermore,
Next to Leading Order as well twist-4 corrections which may be important 
at these rather low $Q^2$ values, still need to be quantified.

This DVCS BSA on the proton has also been measured at higher energies, 
using beams of 4.8 GeV and 5.75 GeV, by the CLAS collaboration and
preliminary results are currently being presented at conferences (see for instance
Ref.~\cite{baryons04franck}).

DVCS cross sections on the proton have, so far, only been measured
at very high energies, by the H1 and ZEUS collaborations~\cite{fav}, 
where gluon contributions dominate. They still lend themselves to GPD 
interpretation through gluon exchange type processes~\cite{freund}
and, again, good agreement is found.

Besides BSA and cross sections, preliminary results for other
observables are regularly presented in conferences~: the 
DVCS beam charge asymmetry (which is sensitive to the real part of the 
amplitude, see section~\ref{thexp}) has been measured on the proton at 
HERMES~\cite{baryons04duren} and the proton target asymmetry at 
CLAS~\cite{baryons04franck}. Also, first results of DVCS on nuclear 
targets are being released~\cite{baryons04duren}.

All these experimental results are very encouraging in the 
sense that the observed signals, although integrated over quite
wide kinematical ranges, are generally compatible (in magnitude 
and in shape) with theoretical calculations. It should be underlined that 
almost all the calculations presented were actual {\it predictions} 
and were published before the experimental results. 
The statistics of all the current measurements are, however, not high enough 
to allow for a fine binning of the kinematical variables and therefore 
do not allow as yet a precise test of the different GPD models.

\subsection{The Prospectives}

In the short-term, new results are soon to be expected from JLab,
using the highest beam energy available.
In Hall A, equipped with high resolution arm-spectrometers,
two dedicated DVCS experiments~\cite{franck,eric} will measure
with high precision some particular kinematical configurations ; the experiment
actually started in fall 2004. In Hall B,
equipped with a large acceptance spectrometer, a DVCS experiment~\cite{volker} 
will allow to cover a broad phase space ; the experiment is scheduled to start 
data taking in spring 2005. These experiments are specialized in the
sense that they will use special additional detectors, designed
to detect the full final state of the reaction $ep\to e^\prime p\gamma$.
This will suppress the background coming from the contamination of additional 
pions. In particular, the Hall B experiment will use
a forward angle electromagnetic calorimeter of about 400 $P_bWO_4$ crystals 
($\approx 1\times \approx 1\times \approx 16$ cm$^3$) equipped with 
Avalanche Photodiodes (APDs) designed to operate in strong magnetic 
fields. They will be used to detect the DVCS photons and unambiguously sign the 
exclusivity of the reaction. A successful one-week run took place in winter 2003
at JLab with a prototype, thus validating this new technique. The forthcoming program 
at JLab also plans to simultaneously measure the DVCS BSA on a deuterium/neutron target.

At HERMES, there is a project~\cite{kaiser,baryons04duren,baryons04franck}   
to install in 2005 a recoil detector for the detection of the recoiling hadronic state.
This should ensure the exclusivity of the reaction.

COMPASS, with a 100 to 200 GeV beam, has the unique feature of potentially
accessing small $x_B$ at sufficiently large $Q^2$ and could also contribute
to the field of GPDs. An experimental project is currently under study~\cite{nicole}.
A dedicated recoil detector would also be installed in order
to detect all the final particles of the DVCS process. Such a program 
could be envisaged to start around 2009.

While an exploratory study of the GPDs is under way at JLab ($E_e$=6 GeV) and
HERMES ($E_e$=27 GeV), and could be envisaged in the future at COMPASS 
($E_\mu$=100-200 GeV), it will probably not be sufficient to fully reveal
the full GPD physics. A lepton accelerator facility 
combining a high luminosity ($\approx 10^{35-36}$ cm$^{-2}$ s$^{-1}$ desirable)
and a high energy beam ($\approx$ 30 GeV) with a good resolution detector 
(a few tens of MeV for a typical missing mass resolution) would be appropriate. 

The JLab upgrade~\cite{cebaf11} with an 11 to 12 GeV beam energy project, planned for 2008, 
promises to be the facility suited to a physics program 
devoted to the systematic study of exclusive reactions and the GPDs.
It meets many of the requirements needed for this ``exclusive" physics which are~:
\begin{itemize}
\item Large kinematical range : it is desirable to span a domain in $Q^2$ and
$x_B$ as large as possible. Although a higher beam energy would be
even better, 11 GeV allows to reach $x_B$ down to 0.1 and $Q^2$ up to 8 GeV$^2$.
\item High luminosity : cross sections fall sharply with $Q^2$ ; for a large acceptance 
detector, luminosities up to $10^{35}$ cm$^{-2}$ s$^{-1}$ are necessary,
\item Good resolution : it is important to cleanly identify \underline{exclusive}
reactions. This can be achieved with a good resolution using the
missing mass technique and/or by detecting \underline{all} the final state particles
and thus overdetermining the kinematics
of the reaction. The CLAS++ upgrade project of the JLab large acceptance 
detector of Hall B is perfectly suited to achieving these goals.
\end{itemize}

\begin{figure}[htb]
\epsfxsize=8. cm
\epsfysize=8. cm
\centerline{\epsffile{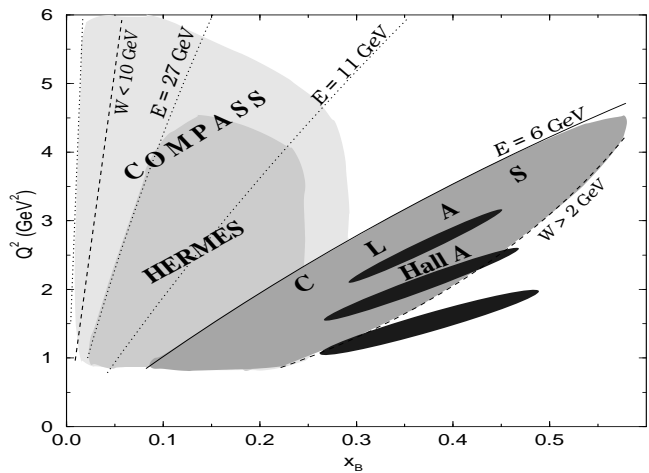}}
\vspace{-1.5cm}
\caption{($Q^2$, $x_B$) phase space covered by the current and short-term 
future experimental projects for DVCS. The figure is taken from Ref.~\cite{volker}.}
\label{fig:comp}
\end{figure}

Figure~\ref{fig:comp} summarizes the phase space covered by the current 
and short-term future projects related to this physics and clearly 
illustrates the complementarity of each.

\section{Conclusion}

In conclusion, we believe that Compton scattering, from the nucleon down 
to the quark and gluon level and its theoretical interpretation in terms 
of GPD, open a broad new area in the investigation of the nucleon structure. 
GPDs allow to access new information in this field such as 
momentum and space correlations between quarks, quarks' orbital momentum and 
quark-antiquark configurations in the nucleon.

The first experimental results on DVCS, from the JLab, HERMES and H1/ZEUS
facilities are encouraging and indicate that
the manifestations of the handbag diagrams, that is to say Compton scattering at the 
quark level, are being seen. Now, dedicated programs at the existing facilities with special
equipment and detectors have started and will soon yield a wealth of new
experimental data. A full study aiming at the extraction of these GPDs from experimental
data probably requires a new devoted facility providing high energy and high
luminosity lepton beams, equipped with large acceptance and high resolution
detectors. JLab with an 11 to 12 GeV beam energy is probably where the future of this 
field lies.

The author wishes to thank M. Vanderhaeghen, B. Pire and M. Mac Cormick
for useful discussions.

\end{document}